\documentclass[aps,twocolumn,tightenlines]{revtex4}
\usepackage[dvipdf]{graphicx}

\begin{document}
\title{Dispersion compensation in atom interferometry by a Sagnac phase}

\author{ Marion Jacquey}
\altaffiliation[present address: ]{Dipartimento di Fisica and
LENS, Universita di Firenze - INFN Sezione di Firenze, via Sansone
1, 50019 Sesto Fiorentino, Italy}

\author{ Alain Miffre}
 \altaffiliation[present address: Lab ]{Universit\'e de Lyon, Universit\'e Lyon 1, CNRS,
 LASIM UMR 5579, b\^at. A. Kastler, 10 rue A. M. Amp\`ere, 69622 Villeurbanne,
 France }

\author{ G\'erard Tr\'enec}
\author{ Matthias B\"uchner}
\author{ Jacques Vigu\'e}
\email{jacques.vigue@irsamc.ups-tlse.fr} \affiliation{ Laboratoire
Collisions Agr\'egats R\'eactivit\'e UMR 5589,
\\ CNRS - Universit\'e de Toulouse-UPS, IRSAMC, Toulouse, France }

\author{ Alexander Cronin}
\affiliation{Department of Physics, University of Arizona, Tucson,
Arizona 85721, USA}

\date{\today}

\begin{abstract}
We reanalyzed our atom interferometer measurement of the electric
polarizability of lithium now accounting for the Sagnac effect due
to Earth rotation. The resulting correction to the polarizability
is very small but the visibility as a function of the applied
phase shift is now better explained. The fact that the Sagnac and
polarizability phase shifts are both proportional to $v^{-1}$,
where $v$ is the atom velocity, suggests that a phase shift of the
Sagnac type could be used as a counterphase to compensate the
electric polarizability phase shift. This exact compensation opens
the way to higher accuracy measurements of atomic polarizabilities
and we discuss how this can be practically done and the final
limitations of the proposed technique.

\end{abstract}
\maketitle

\section{Introduction}

We have measured the electric polarizability of lithium atoms
\cite{miffre06a,miffre06b} by atom interferometry, with an
experiment very similar to the one of C. R. Ekstrom {\it et al.}
on sodium \cite{ekstrom95}. The measured phase shift as a function
of the applied voltage was well fitted by a theoretical analysis
in which the Sagnac phase shift due to the Earth rotation was
neglected. As these two phase shifts are both dependent on the
atom velocity, this omission has an effect on the predicted fringe
visibility. In the present paper, we reanalyze our data with the
Sagnac phase taken into account. The correction to the
polarizability value is very small but the new fit corrects
substantially the best fit width of the velocity distribution of
the atomic beam.

This result would have only a minor interest if it does not
suggest a way of improving polarizability measurements by atom
interferometry. As discussed by T. D. Roberts {\it et al.}
\cite{roberts04}, the main factor limiting the accuracy of this
measurement comes from the dependence of the polarizability phase
shift $\Delta\phi_{pol}$ with the atom velocity $v$
($\Delta\phi_{pol}\propto v^{-1}$) and this paper developed an
experiment in which the polarizability phase shift was compensated
by a velocity dependent counterphase. In this way, it was possible
to increase the maximum observable phase shift, with the goal of
improving the measurement accuracy. The counterphase used in
reference \cite{roberts04} was not exactly proportional to
$v^{-1}$ and this defect has probably limited the performance of
this technique. We nevertheless think that the idea of a velocity
dependent counterphase is excellent and we propose to use the
Sagnac phase shift
\cite{overhauser74,collela75,anandan77,werner79,riehle91} as a
counterphase. As this phase shift is exactly proportional to
$v^{-1}$, it should provide an ideal compensation. To produce a
Sagnac phase shift, it is possible to rotate the whole atom
interferometer, as done by Lenef {\it et al.} \cite{lenef97}, but
more convenient techniques can be used. For example, when the atom
wave is diffracted by lasers, moving only two small mirrors in
opposite directions would have the same effect. In the case of
Raman diffraction, a phase shift proportional to $v^{-1}$ can be
created by varying the Raman frequency shift and such a phase
shift has already been used in gyrometers based on atom
interferometry \cite{gustavson97,gustavson00,canuel06}. We thus
think that we can develop a very accurate counterphase which will
be useful for high-accuracy measurements of atom electric
polarizability.

\section{Electric polarizability measurement by atom interferometry}

\begin{figure}
\includegraphics[width = 8 cm,height= 6.0 cm]{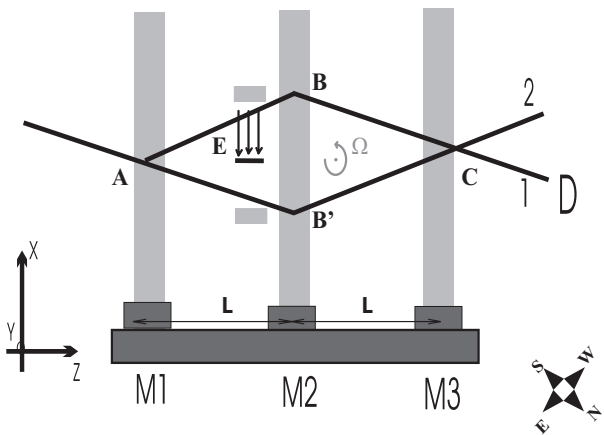}\\
\caption{Our measurement of an electric polarizability with a
Mach-Zehnder atom interferometer: a collimated supersonic lithium
beam is diffracted, in the Bragg regime, by three laser standing
waves produced by reflecting three laser beams on the mirrors
$M_1$, $M_2$ and $M_3$.  The output beam labelled $1$ is selected
by a slit and detected by surface ionization on a hot wire
detector $D$. A capacitor with a thin electrode is inserted
between the two interfering atomic beams, just before the second
laser standing wave, where the distance between the two atomic
beams is largest. The electric field $E$ is applied on the upper
path of the interferometer. The orientation of the experiment is
roughly represented as well as the sense of the Earth angular
velocity component $\Omega_y$.} \label{interferometer}
\end{figure}

The principle of an electric polarizability measurement by atom
interferometry \cite{ekstrom95} is to use a Mach-Zehnder atom
interferometer in which the atomic beams can be separated by a
septum and to apply an electric field on only one of the two
atomic beams (see figure \ref{interferometer}). For a given
voltage $U$ applied on the capacitor creating the field, the
polarizability phase shift $\Delta\phi_{pol}(U)$ of the atomic
wave is given by:

\begin{equation}
\label{n1} \Delta\phi_{pol}(U) = \frac{2 \pi \epsilon_0 \alpha
}{\hbar v} \int E^2(z) dz
\end{equation}

\noindent Here $v$ is the atom velocity $v = \hbar k/m$ (for more
details, see references \cite{miffre06a,miffre06b}). An important
point here is that the polarizability phase shift
$\Delta\phi_{pol}(U)$ is proportional to $v^{-1}$ and we will write

\begin{equation}
\label{n1a} \Delta\phi_{pol}(U) = \Delta\phi_{pol,m}(U) u/v
\end{equation}

\noindent where $\Delta\phi_{pol,m}$ is the phase shift for the
atomic beam mean velocity $u$ defined below. We may note that the
proportionality $\Delta\phi_{pol}(U) \propto v^{-1}$ is the result
of a first order perturbation calculation, valid if the ratio of
the polarizability energy term to the atom kinetic energy (i.e.
the ratio $2 \pi \epsilon_0 \alpha E^2(z) /\left(\hbar^2
k^2/2m\right)$) is considerably smaller than unity. In the
experiments with thermal atoms
\cite{ekstrom95,miffre06a,miffre06b}, this ratio is typically of
the order of $10^{-8}$ and this approximation is excellent.

Finally, the velocity dependence of the phase shift coupled to the
velocity dispersion of the atomic beam induces a rapid reduction of
the fringe visibility when the applied phase shift increases: as
discussed in \cite{roberts04}, this effect limits the maximum
measurable phase shift and ultimately the precision of the
polarizability measurement.

\section{Sagnac effect}

Atom interferometers are extremely sensitive to inertial effects
and in particular to rotation of the setup, through Sagnac effect
\cite{overhauser74,collela75,anandan77,werner79,riehle91}. Several
gyrometers based on atom interferometry have been developed
\cite{riehle91,lenef97,gustavson97,gustavson00,canuel06} and an
extremely high sensitivity has been achieved \cite{gustavson00}.
The Sagnac phase shift due to a rotation of the setup is given by:

\begin{equation}\label{n2}
 \Delta \phi_{Sagnac}= 2 k_G \Omega_y L^2/v
\end{equation}

\noindent Here, $k_G$ is the grating wave vector ($k_G= 2 k_L$ in
the case of laser diffraction by a laser with a wave vector
$k_L$);  $\Omega_y$ is the $y$-component of the angular velocity
of the laboratory frame with respect to a Galilean frame (the
$y$-axis being perpendicular to the plane of the atom trajectory,
see figure \ref{interferometer}); finally $L$ is the distance
between consecutive gratings. In our experiment, the mean value of
$\Omega_y$ is due to the Earth rotation, while the seismic and
laboratory vibrations induce rapid fluctuations of $\Omega_y$. The
main effect of these fluctuations is to induce a phase noise which
reduces the fringe visibility \cite{miffre06,jacquey 06}. With a
period equal to the sidereal day, $\Omega_y = 5.025 \times
10^{-5}$ rad/s at the laboratory latitude
$\lambda=43^{\circ}33'37''$, $k_L = 9.364\times 10^6$ m$^{-1}$ and
$L=0.605$ m, the calculated Sagnac phase shift is given by:

\begin{equation}\label{n3}
\Delta \phi_{Sagnac}=0.646\times u/v \mbox{ rad}
\end{equation}
\noindent Here $u $ is the beam mean velocity defined by equation
(\ref{n5}) below, and $u =1065.7 \pm 5.8 $ m/s as measured in
\cite{miffre06a,miffre06b}.

\section{Reanalysis of our experimental signals}

As the phase shifts $\Delta\phi_{pol}$ and $\Delta \phi_{Sagnac}$
are velocity dependent, both in $v^{-1}$, we must describe the
velocity distribution of the atomic beam. As in references
\cite{miffre06a,miffre06b}, we will use:

\begin{equation}
\label{n5} P(v)  = \frac{S_{\|}}{u \sqrt{\pi}} \exp\left[-\left(
(v-u)S_{\|}/u\right)^2\right]
\end{equation}

\noindent This equation is a simplified form of the velocity
distribution of supersonic beams \cite{beijerinck83,haberland85},
with $S_{\|}$, the parallel speed ratio, defined such that $S_{\|}
= u / (\sigma\sqrt{2})$ where $\sigma$ is the RMS about the mean
velocity. In fact, $P(v)$ is not the velocity distribution of the
incident beam but the velocity distribution of the atoms
contributing to the fringe signal i.e. it is the product of the
velocity distribution of the incident beam by the transmission of
the interferometer. This transmission is a function of the atom
velocity, in particular because of the use of Bragg diffraction.
Noting $\Delta\phi = \Delta\phi_{pol}(U) + \Delta \phi_{Sagnac}$,
the interferometer signal is given by:

\begin{eqnarray}\label{n4}
I &=& I_0 \int dv P(v)  \left[1 + {\mathcal{V}}_0
\cos\left( \psi + \Delta \phi  \right) \right] \nonumber \\
&=&I_0\left[ 1+\langle \mathcal{V}\rangle \cos(\psi+\langle
\Delta\phi \rangle)\right]
\end{eqnarray}

\noindent $\psi$ is a phase shift function of the standing wave
positions, which serves to observe interference fringes and which
is independent of the atom velocity. Equation (\ref{n4}) defines
the observed visibility $\langle \mathcal{V}\rangle$ and the
observed phase shift $\langle \Delta\phi \rangle$. These two
averages are non linear, so that:

\begin{equation}\label{n6}
\langle \Delta\phi \rangle \neq \int dv P(v)  \Delta \phi (v)
\end{equation}
\noindent As a consequence, after taking the average, the
polarizability and Sagnac phase shifts are not exactly additive.
To measure the effect of an applied voltage $U$, we make two
measurements of the fringe phase, one with $U=0$ and one with $U$
and the difference provides our measurement of the polarizability
phase shift $\left(\Delta\phi_{pol}(U)\right)_{meas}$ given by:

\begin{equation}\label{n7}
\left(\Delta\phi_{pol}(U)\right)_{meas} = \langle
\Delta\phi_{pol}(U) + \Delta \phi_{Sagnac} \rangle - \langle
\Delta \phi_{Sagnac} \rangle
\end{equation}

\noindent As the electric field was applied on the ABC beam of the
interferometer (see figure \ref{interferometer}), one can verify
that the polarizability and Sagnac phase shifts have opposite
signs. We use equations (\ref{n5},\ref{n4},\ref{n7}) to fit our
data with only two adjustable parameters, the parallel speed ratio
$S_{\|}$ and the ratio $\Delta\phi_{pol,m}(U)/U^2$. The phase
shift and visibility data are treated in a single fit and we get:

\begin{eqnarray}\label{n8}
S_{\|}&=&7.67\pm0.06\\
\frac{\Delta\phi_{pol,m}(U)}{U^2}&=&(1.3880\pm0.0010)\times10^{-4}\textrm{rad/V}^2
\end{eqnarray}

\noindent In our initial fit neglecting 
the Sagnac effect \cite{miffre06a,miffre06b}, the parallel speed
ratio $S_{\|}$ was found equal to $ S_{\|} =  8.00 \pm 0.06$ and
this overestimation can be explained: the Sagnac and polarizability
phase shifts being of opposite signs, the visibility decay is
delayed by the existence of the Sagnac phase shift. This delay
induced the initial fit toward a velocity dispersion smaller than
its actual value.

The $\Delta\phi_{pol,m}(U)/U^2$ value is only slightly modified
with respect to its previous value equal to
$\Delta\phi_{pol,m}(U)/U^2 = (1.3870\pm 0.0010) \times 10^{-4}
\mbox{ rad/V}^2$. This is not surprising because the effect is
very indirect, through the non-linear character of the average
defined by equation (\ref{n4}). Our new value of the
polarizability of $^7$Li is:

\begin{equation}\label{n9}
\alpha(Li)=(24.34\pm0.16)\times10^{-30} \mbox{ m}^3=164.3\pm1.1
\mbox{ u.a.}
\end{equation}

\begin{figure}
\includegraphics[width = 8 cm,height= 7 cm]{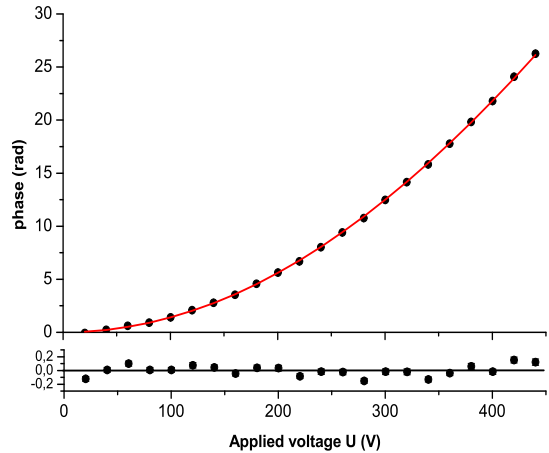}
\caption{(Color online) Measured phase shift
$\Delta\phi_{pol,m}(U)$ in radians as a function of the applied
voltage $U$: the best fit is represented by the full curve and the
residuals are plotted in the lower graph with an expanded scale.
}\label{fig2}
\end{figure}

\begin{figure}[h!]
\includegraphics[width = 8 cm,height= 6.0 cm]{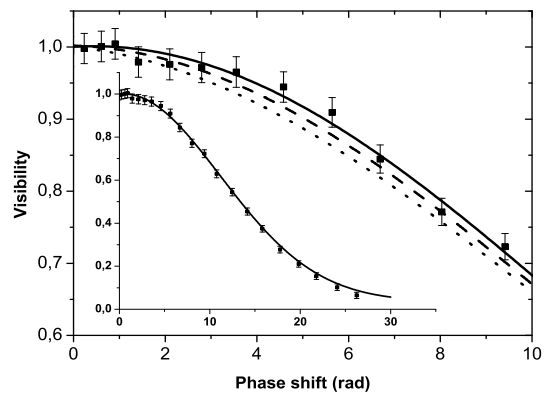}
\caption{ The reduced fringe visibility $\langle
\mathcal{V}(U)\rangle/ \langle \mathcal{V}(U=0)\rangle$ is plotted
as a function of the measured phase shift $\Delta\phi_{pol,m}(U)$
(the zero-field visibility is equal to $\langle
\mathcal{V}(U=0)\rangle = 62$ \%). The data points are plotted
with their error bars and three different curves are plotted: with
the correct Sagnac phase shift (full curve), without the Sagnac
phase shift (dashed curve) and finally with the Sagnac phase shift
taken into account with the wrong sign (dotted curve). The insert
shows the new fit taking into account the Sagnac term over our
complete data set.} \label{fig3}
\end{figure}

\noindent The $0.07$ \% correction is considerably smaller than
the $0.66$ \% final error bar dominated by the uncertainty on the
mean velocity $u$. Figures \ref{fig2} and \ref{fig3} show the
agreement between the data points and their fits. On the
visibility data, the improvement is noticeable, especially when
the measured phase shifts are smaller than $10$ rad.


\section{The use of a Sagnac counterphase for dispersion compensation}

The fact that the polarizability phase shift is proportional to
$v^{-1}$ limits the accuracy of the measurement for three main
reasons:

\begin{itemize}

\item the visibility decreases rapidly when the mean phase shift
increases and for sufficiently large phase shifts, the phase
sensitivity is too small to be useful. T. D. Roberts {\it et al.}
\cite{roberts04} have made a quantitative analysis of this effect,
with an evaluation of the optimum phase shift in the case of a
$v^{-n}$ dependence of the phase shift with the velocity $v$.

\item as shown by equations (\ref{n4},\ref{n7}), the measured
phase shift $\left(\Delta\phi_{pol}(U)\right)_{meas}$ differs from
the phase shift $\Delta\phi_{pol,m}(U)$ corresponding to the mean
velocity $u$ (see also Appendix B of ref. \cite{miffre06b}). This
effect is very important when the velocity distribution is broad
and it forbids a very accurate determination of the
polarizability.

\item finally, the mean velocity $u$ is difficult to determine
with great accuracy, especially because the incident beam mean
velocity is slightly modified by the velocity dependent
transmission of the interferometer.

\end{itemize}

T. D. Roberts {\it et al.} proposed to compensate the
polarizability phase shift by an engineered counterphase. To
produce this counterphase, two time dependent phase shifts
$\phi_{1,2} (t)$ were applied at two points separated by a
distance $L_{shifters}$ along the atom path in the interferometer.
The counterphase $\phi_{counter}$ is thus given by:

\begin{equation}
\label{n10} \phi_{counter} = \phi_{1} \left(t\right) + \phi_{2}
\left( t+ L_{shifters}/v \right)
\end{equation}
\noindent We will not reproduce the discussion of T. D. Roberts
{\it et al.} to which we refer the reader. However, we point out,
that in addition to the velocity dependence due to the
$L_{shifters}/v$ term, there is a direct velocity dependence in
$v^{-2}$ of $\phi_{1}$ and $\phi_{2}$ due to the fact that these
phase shifts are produced by applying an electric field gradient
on the atomic beams inside the interferometer. This supplementary
velocity dependence, which was not discussed in reference
\cite{roberts04}, complicates the use of this counterphase, and we
note that a high accuracy use of this technique remains to be
demonstrated

\section{The use of Sagnac phase shift for dispersion compensation}

The observation presented in this paper suggests that one may use
the Sagnac phase shift as a counterphase to compensate the
polarisability phase shift. As both phase shifts are proportional
to $v^{-1}$, the compensation should be exact and the only problem
is to find a practical way of creating a large Sagnac phase shift.

If we move the mirrors of the first and/or third standing waves in
opposite directions, these motions will exactly mimick a rotation
of the interferometer. If the velocity of the mirrors $M_1$ and
$M_3$ are respectively $v_1$ and $v_3$, the induced phase shift is
given by \cite{schmiedmayer97} :

\begin{equation}\label{n11}
 \Delta \phi_{Sagnac}= 2 k_L (v_1- v_3)L/v
\end{equation}

\noindent This idea seems nice, especially as piezo-actuators with
capacitive displacement sensors are available, with an uncertainty
on the displacement smaller than $1$ nanometer. For a phase shift
$\Delta \phi_{Sagnac}= 100$ radians, the needed velocities are
quite large $v_1=-v_3 = 4.7 \times 10^{-3}$ m/s, which can be
sustained only during about $4\times 10^{-3}$ s if the
displacement is limited to $20$ micrometers. The mirrors should be
moved forward and backward with a few millisecond period. This
rapid motion will perturb the interferometer and we think that
this technique is not practically feasible.

\begin{figure}
\includegraphics[width = 7 cm,height= 6.0 cm]{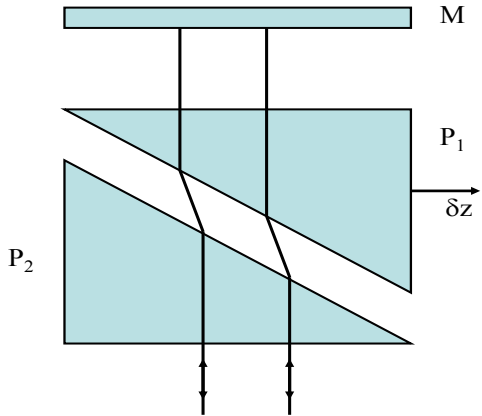}\\
\caption{(Color online) Two-prism arrangement proposed to create a
Sagnac-type phase shift. The two prisms $P_1$ and $P_2$ are
identical and they are introduced on the laser standing wave
between the atomic beams and the mirror $M$. Their angle is chosen
such that the laser beam can be at normal on one face and at
Brewster incidence on the other face. The displacement $\delta z$
of prism $P_1$ induces a modification $\delta x$ of the one-way
optical path for the laser, which is equivalent to a $\delta x$
displacement of the mirror perpendicular to its surface}
\label{prismSagnac}
\end{figure}

However, what we need is to change the positions of the nodes and
anti-nodes of the laser standing waves and a two-prism arrangement
as schematically represented in figure \ref{prismSagnac} can do
the same effect. A displacement $\delta z$ of the prism is
equivalent to a change of the mirror position $\delta x$ given by:

\begin{equation}\label{n12}
\delta x =  \delta z \frac{1-n \cos(i-r)}{n} = \delta z
\frac{1-n^2}{n(1+n^2)}
\end{equation}

\noindent where $i$ and $r$ are the incidence and refraction
angles of the laser beam on the prism hypotenuse. The final form
assumes Brewster incidence as drawn on figure \ref{prismSagnac}
(i.e. $\tan i = n$ where $n$ is the index of refraction of the
prism). With fused silica ($n\approx1.46$), the ratio $\delta x/
\delta z$ is close to $-0.25$ and such an arrangement can replace
a mirror displacement of several millimeters. It thus seems
feasible to produce a Sagnac type phase shift of the order of
$100$ radians during a time period of the order of $1$ second.
Commercially available translation stages with interferometric
control of the position can provide a very stable prism velocity
which is needed because any velocity fluctuation will induce a
phase noise.

The prism motion will induce a large Sagnac counterphase $\Delta
\phi_{Sagnac}$ and the measurement will give $\left(\Delta
\phi_{tot}\right)_{meas} = \langle \Delta\phi_{pol}(U)\rangle
+\langle \Delta \phi_{Sagnac}\rangle$. We must tune the Sagnac
phase till $\left(\Delta \phi_{tot}\right)_{meas}$ is very small
and then its value is, with a good approximation, the value for
the mean velocity $u$ :

\begin{equation}
\frac{2 \pi \epsilon_0 \alpha }{\hbar} \int E^2(z) dz =  2 k_L
(v_1- v_3)L + u \left(\Delta \phi_{tot}\right)_{meas}
\end{equation}

\noindent If the phase $\langle \Delta\phi \rangle $ is very
small, a low accuracy on $u$ is sufficient for an accurate
measurement of the polarizability $\alpha$.  This measurement will
require an accurate measurement of the quantity $2 k_L (v_1-
v_3)L$. The laser wave vector $k_L$ is easily measured with a very
high accuracy and it is possible to measure the velocity $v_1$
and/or $v_3$ with a high accuracy by optical interferometry (the
laser standing waves used for diffraction can be used in Michelson
interferometers for this measurement). The distance $L$ between
laser standing waves may be difficult to measure with an accuracy
better than about $10^{-3}$, because, as shown by C.J. Bord\'e
\cite{borde04} and Ch. Antoine \cite{antoine04,antoine06}, the
exact value of the distance $L$ is not exactly equal to the
physical distance between the laser standing waves but its value
depends on the diffraction regime and laser beam parameters.

Another technique to create a Sagnac type counter phase is to use
laser Raman diffraction. This diffraction process, which is most
commonly used in cold atom interferometers, has also been used by
T. L. Gustavson {\it et al.} in their thermal atom gyrometer
\cite{gustavson97,gustavson00}. With Raman diffraction, the phase
of the interferometer signal can be modified by changing the
frequency difference of the Raman laser beams. A Sagnac phase can
be mimicked by applying opposite frequency offsets to the Raman
beams used for the first and third diffraction events. T. L.
Gustavson {\it et al.} \cite{gustavson00} have shown that this
phase shift is proportional to $v^{-1}$ and they used it to
compensate the Sagnac phase shift. Such an electronic compensation
can be used to compensate a very large polarizability phase shift.

\section{Conclusion}

In this paper, we have reanalyzed our measurement of the electric
polarizability of lithium atom by atom interferometry and this
reanalysis has taken into account the Sagnac phase shift due to
Earth rotation, an effect which was ignored in our first analysis.
The new value of the lithium electric polarizability differs only
slightly (by $0.07$ \%) from our previous value, well within our
$0.66$ \% error bar. However, the fit to the visibility data is
improved and the deduced value of the parallel speed ratio
$S_{\|}$ is noticeably different.

The polarizability phase shift and the Sagnac phase shift have the
same $v^{-1}$ dependence with the atom velocity $v$ and this
remark suggests that the Sagnac phase shift can be used to
compensate the polarizability phase shift. The idea of
compensating the polarizability phase shift by a counterphase was
initially developed and demonstrated by T. D. Roberts {\it et al.}
\cite{roberts04} and this technique should give access to high
accuracy measurements of electric polarizability. In the
experiment of T. D. Roberts {\it et al.}, the counterphase had not
exactly the same velocity dependence and this effect limits the
possible accuracy. With a Sagnac phase shift as a counterphase,
the compensation should be exact and the polarizability
measurement will be replaced by a measurement of mirror velocity.
Furthermore, if the interferometer uses Raman diffraction, the
Sagnac phase can be mimicked by laser frequency offsets, which are
even simpler to measure. In both cases, the size $L$ of the
interferometer plays a role and its exact value will depend on the
diffraction regime. The test of these compensation schemes require
substantial modifications of our atom interferometer so that they
cannot be done immediately but we expect to do so in a near
future.

\section{Acknowledgements}

The Toulouse group thanks CNRS department MPPU for its support,
R\'egion Midi-Pyr\'en\'ees for a 2005-2006 PACA-MIP network and
ANR for Grant ANR-05-BLAN-0094. AC thanks NSF for Grant No.
PHY-0653623.


\end{document}